\documentclass[11pt]{article}
\begin{document}\hbadness=10000
\pagestyle{myheadings}\markboth{Ludwik Turko,  Jan Rafelski}
{Dynamics of Multiparticle Systems with non - Abelian Symmetry}
\title{Dynamics of Multiparticle Systems with non -- Abelian
Symmetry}
\author{
$\ $\\
\bf  Ludwik Turko\footnote{E-mail: turko@ift.uni.wroc.pl}$^{\ 1,2}$ and
Jan Rafelski\footnote{E-mail: rafelski@physics.arizona.edu}$^{\ 1}$\\
$\ $\\
$^1$Department of Physics, University of Arizona\\
 Tucson, AZ 85721, U.S.A.\\
$\ $\\ and\\ $\ $\\
$^{2}$Institute of Theoretical Physics, University of Wroclaw\\
pl. Maksa Borna 9, 50-204 Wroclaw, Poland}
\date{March 6, 2000}   
\maketitle
\thispagestyle{empty}
\begin{abstract}
{\noindent
We consider the dynamics governing the evolution of
a many body  system constrained by an
nonabelian local  symmetry. We obtain explicit
forms of the global macroscopic
condition assuring that at the  microscopic level  the evolution
 respects the overall symmetry constraint. We demonstrate the
constraint mechanisms for the case of $SU(2)$ system comprising
particles in fundamental, and adjoint representations (`nucleons'
and `pions').\\

PACS: 11.30.-j, 05.20.Dd, 12.38.Mh, 25.75. -q
 }
\end{abstract}
\section{\normalsize\bf Introduction and Overview}\label{intr}
\noindent
 The consideration of the influence of internal symmetries on the
final state of a many body system begun with the pioneering work
of Bethe \cite{Bethe}. Much of the subsequent interest in the
subject arises from the realization that in the study of hadronic
interactions and in particular in studies involving quark
confinement, these constraints may be of decisive importance. An
important progress in treating equilibrium systems was made
employing group projection techniques. This allowed for a
consistent treatment of abelian \cite{Magal} and nonabelian
\cite{Cerul,Erics,Zalew,Erik,Skag} symmetries of compact groups
and a consistent formulation of thermodynamics of many particle
systems with internal symmetries taken into account
\cite{TurRed,Turko}. Application of these methods to specific
processes demonstrated in which circumstances the presence of
symmetry is of physical relevance \cite{RafDan,MullRaf,Elze}.

However, it is not fully understood how the symmetry-modified
properties of the  equilibrium system arise from kinetic
formulation of the dynamical evolution.  When an internal symmetry
is not at work, Boltzmann's H theorem in principle assures that
the statistical Bose/Fermi/Boltzmann distributions are the
asymptotic (equilibrium) distributions, irrespective of the nature
of microscopic interaction. However, in presence of exact
symmetries the equilibrium distributions are modified, see e.g.
\cite{Turko,MullRaf}.  This implies that symmetry constraints
introduce effective interactions of potentially far more complex
nature than is the usual two body Boltzmann collision term. In
fact it can be argued that quantum symmetry constraints are the
heart of the nonlocality of quantum physics. However, in the limit
of classical Boltzmann equation evolution these are implemented by
a strictly local (though non-linear) consideration of Fermi
blocking and Bose enhancement in phase space evolution.  Our aim
in this work is to make a step towards understanding how the
microscopic  nonabelian symmetry constrains operate within the
kinetic master equation description of the time evolution, leading
on to the symmetry modified (constrained) macroscopic many
particle equilibrium state.

It is first important to convince oneself that an underlying
symmetry of microscopic interactions does not lead in general to
the desired symmetry properties of a (macroscopic) many body
interacting  system. To do this, we consider the high energy
nuclear (heavy ion) collisions and specifically here two
symmetry examples:\\

a) {\it SU(2) Isospin symmetry}: The initial state transforms
under a given representation of the isospin $SU(2)$ group. All
elementary high energy interactions are governed by the strong
interaction, which preserves the isotopic symmetry. A final state
results as a multiparticle state formed by many individual hadron
-- hadron collisions. In any of such microprocesses the isospin is
conserved. However, proceeding `as usual' without symmetry
constrained treatment of local interactions  does not assure that
the final multiparticle state (macrostate)  transforms under the
same representation of the isospin group as the initial state,
which is required for
symmetry reasons. \\

b) {\it SU(3) Colour symmetry}: A similar situation appears in the
context of the quark-gluon interactions, especially in case that
local deconfinement occurs. The initial state is a colour singlet
state, and  quark-gluon and gluon-gluon interaction, although
invariant under the colour $SU(3)_c$ symmetry group do not assure
that during its evolution a (macroscopic) many particle state,
once a singlet, always remains a singlet colour state, which,
however, it must do because
of exact colour symmetry of strong interactions.\\

As these examples show, in a dynamical (quantum) transport theory
description of the approach to equilibrium there must exist a
subsidiary condition which should be taken into account by
corresponding kinetic equations governing the evolution. This
condition is independent from other constraints related to
dynamical gauged internal symmetries. For classical fluid dynamics
a dynamical evolution equation addressing gauge symmetry has been
proposed by Wong \cite{Won70}. In many current studies of the
dynamics of classical non-abelian fields this proposal continues
till today to attract considerable interest \cite{LM99}. However,
these are constraints which have no relation to the intrinsic
non-locality of the quantum system which we address here.

We first note that in the case of an abelian symmetry there are no
additional constraints to consider. Quantum number conservation on
a microscopic level is fully equivalent to preservation of all
symmetry properties on the macroscopic level. This is easily seen
considering the $U(1)$ symmetry related to microscopic
particle-antiparticle formation: since the microscopic mechanisms
produce equal number of particles and antiparticles (pair
production), initial particle-antiparticle number difference is
exactly preserved in the  macroscopic many body state.

Thus only presence of a nonabelian symmetry poses a true
challenge. A suitable mathematical method how to approach this
problem is identified considering  the previously treated
statistical equilibrium case.  Here one decomposes a general
`macrostate' consisting of many particles into possible
irreducible representations of the symmetry group. Then a
projection technique exploiting character function properties of
the group is used to constrain the final state. We will here use
this approach in order to describe multiparticle evolution
applying microscopic kinetic theory scheme. We will show that the
behaviour of particle phase - space distribution functions will
depend not only on properties of basic interaction, but that it
also depends on global properties of the macroscopic system. Those
global properties provide subsidiary constraints needed, so that
the asymptotic equilibrium state has properties  consistent with
the non-abelian constraints.
\section{\normalsize\bf The Projection Method}\label{proj}
\noindent
 Let $G$ be a compact internal symmetry group of our system
consisting  of particles (objects) transforming under irreducible
representations of the symmetry group. These representation are
denoted as $\alpha_i$ with corresponding dimensions $d(\alpha_i)$.
One denotes $f^{(\alpha_i,\nu_i)}_{(\zeta)}(\Gamma,\vec r,t)$ a
distribution function of the particle which belongs to the
multiplet $\alpha_i$ of the symmetry group. Members of this
multiplet are numbered by indexes $\nu_i\
(\nu_i=1,\dots,d(\alpha_i))$ which correspond to given values of
charges related to the symmetry group. A subscript $\zeta$ denotes
other quantum numbers characterizing different multiplets of the
same representation $\alpha$.  The variables $(\Gamma,\vec r)$
denotes a  set of the phase - space variables such as $(\vec
p,\vec r)$ and $t$ is time.

The number of particles of the specie
$\{\alpha,\nu_\alpha,\zeta\}$ is:
\begin{equation}
N^{(\alpha)}_{\nu_\alpha;(\zeta)}(t)=\int\,dV d\Gamma
f^{(\alpha,\nu_\alpha)}_{(\zeta)}(\Gamma,\vec r,t)\,; \label{numb}
\end{equation}
We consider a system of
 $\{N_{\alpha_1, \nu_{\alpha_1}}^{(\zeta_1)}(t),\dots,
{N_{\alpha_n,\nu_{\alpha_n}}^{(\zeta_n)}(t)}\}$ particles at time
$t$\,.  The distribution functions fulfill the generalized Vlasov
- Boltzmann kinetic equations, which can be written in the general
form:

\begin{eqnarray}\label{kin}
{\partial f^{(\alpha_i,\nu_i)}_{(\zeta_i)}(\Gamma_i,\vec r,
t)\over\partial t}& + & \vec v\cdot\nabla
f^{(\alpha_i,\nu_i)}_{(\zeta_i)}(\Gamma_i,\vec r,t)\nonumber\\
&=&\sum_{\alpha_j,\alpha_k,\alpha_l}\;\sum_{\nu_j,\nu_k,\nu_l}\;
\sum_{\zeta_j,\zeta_k,\zeta_l}\int\, d\Gamma_j d\Gamma_k
d\Gamma_l\, {\cal
W}_{\nu_i\nu_j;\nu_k\nu_l}^{(\zeta_i,\zeta_j;\zeta_k,\zeta_l)}
(\Gamma_k,\Gamma_l;\Gamma_j,\Gamma_i)\nonumber\\ &\ &\left[ {\cal
F}^{(\alpha_i,\nu_i)}_{(\zeta_i)}(\Gamma_i,\vec r,t){\cal
F}^{(\alpha_j,\nu_j)}_{(\zeta_j)}(\Gamma_j,\vec r,t)
f^{(\alpha_k,\nu_k)}_{(\zeta_k)} (\Gamma_k,\vec
r,t)f^{(\alpha_l,\nu_l)}_{(\zeta_l)}(\Gamma_l,\vec r,t)\right.
\nonumber\\ &&\left.-{\cal
F}^{(\alpha_k,\nu_k)}_{(\zeta_k)}(\Gamma_k,\vec r,t){\cal
F}^{(\alpha_l,\nu_l)}_{(\zeta_l)}(\Gamma_l,\vec r,t)
f^{(\alpha_i,\nu_i)}_{(\zeta_i)}(\Gamma_i,\vec
r,t)f^{(\alpha_j,\nu_j)}_{(\zeta_j)} (\Gamma_j,\vec r,t)\right]\,;
\end{eqnarray}

Factors ${\cal F}^{(\alpha,\nu)}_{(\zeta)}(\Gamma,\vec r,t)$ are
related to quantum statistics and they are equal to $1$ for
classical particles, and equal to $[1\pm
f^{(\alpha,\nu)}_{(\zeta)}(\Gamma,\vec r,t)]$ for bosons/fermions
correspondingly. Since by assumption the whole system transforms
under given representation $\Lambda$ of an exact symmetry group,
the system under consideration must preserve its transformations
properties during its time evolution, provided that it is governed
by a symmetry invariant interaction.

We now focus on the case of a quantum system and consider state
vectors in particle number representation: $\left\vert
N^{(\alpha_1)}_{\nu_{\alpha_1}},\dots,
N^{(\alpha_n)}_{\nu_{\alpha_n}}\right\rangle .$ These vectors
describe symmetry properties of our systems and all other
variables, related to phase-space properties of the system are
suppressed here. They transform as a direct product representation
of the symmetry group $G$. This representation is of the form:
\begin{equation}
\alpha_1^{N^{(\alpha_1)}} \otimes \alpha_2^{N^{(\alpha_2)}}
\otimes \cdots \otimes \alpha_n^{N^{(\alpha_n)}}\,; \label{prod}
\end{equation}
A multiplicity $N^{(\alpha_j)}$ of the representation $\alpha_j$
in this product is equal to a number of particles which transform
under this representation:
\begin{equation}
N^{(\alpha_j)} = \sum_j\left(\sum_{\zeta_j}\,
N^{(\alpha_j)}_{\nu_{\alpha_j};(\zeta_j)}\right)= \sum_j\,
N^{(\alpha_j)}_{\nu_{\alpha_j}}\,; \label{pnumber}
\end{equation}
The representation given by Eq. (\ref{prod}) can be decomposed
into direct sum of irreducible representations $\Lambda_k$.
Corresponding states are denoted as $\left\vert \Lambda_k,
\lambda_{\Lambda_k}; {\cal N}\right\rangle$ where
$\lambda_{\Lambda_k}$ is an index numbering members of the
representation $\Lambda$ and ${\cal N}$ is a total number of
particles
\begin{equation}
{\cal N}=\sum_k\, N^{(\alpha_k)}_{\nu_{\alpha_k}}\,;\label{number}
\end{equation}
Each physical state can be decomposed into irreducible
representation base states with amplitudes depending on phase
space variables $\Gamma$:
\begin{eqnarray} \label{dsum}
\left\vert N^{(\alpha_1)}_{\nu_{\alpha_1}},\dots, N^{(\alpha_n)}_{
\nu_{\alpha_n}};\Gamma\right\rangle=
\sum_{k}\!^{\oplus}\sum_{\xi_{\Lambda_k}}\!^{\oplus} \left\vert
\Lambda_k, \lambda_{\Lambda_k}; {{\cal
N};\xi_{\Lambda_k}}\right\rangle
a^{\Lambda,\lambda_{\Lambda}}_{\{N^{(\alpha_1)}_{\nu_{\alpha_1}},
\dots,\,N^{(\alpha_n)}_{\nu_{\alpha_n}}\}}(\xi_{\Lambda_k};\Gamma)\,;
\end{eqnarray}
Here appear new variables $\xi_{\Lambda}$ which are degeneracy
parameters required for the full description of a state in the
"symmetry space".

Let us define an average weight
\begin{equation}\label{avweight}
\overline{P^{\Lambda,\lambda_{\Lambda}}_{\{N^{(\alpha_1)}_{\nu_{\alpha_1}},
\dots,\,N^{(\alpha_n)}_{\nu_{\alpha_n}}\}}}\;=\;
\frac{\sum\limits_{\xi_\Lambda}\vert a^{\Lambda,
\lambda_{\Lambda}}_{\{N^{(\alpha_1)}_{\nu_{\alpha_1}},
\dots,\,N^{(\alpha_n)}_{\nu_{\alpha_n}}\}}(\xi_{\Lambda};\Gamma)\vert^2}
{\sum\limits_{N^{(\alpha_1)}_{\nu_{\alpha_1}}+
\cdots+\,N^{(\alpha_n)}_{\nu_{\alpha_n}}={\cal N}}\;
\sum\limits_{\xi_\Lambda}\vert
a^{\Lambda,\lambda_{\Lambda}}_{\{N^{(\alpha_1)}_{\nu_{\alpha_1}},
\dots,\,N^{(\alpha_n)}_{\nu_{\alpha_n}}\}}(\xi_{\Lambda};\Gamma)\vert^2}\,;
\end{equation}

This expression gives the probability that
$N^{(\alpha_1)}_{\nu_{\alpha_1}},\dots, N^{(\alpha_n)}_{
\nu_{\alpha_n}}$ particles transforming under the symmetry group
representations $\alpha_1,\dots,\alpha_n$ combine into $\cal N$
particle state transforming under representation $\Lambda$ of the
symmetry group.

We make the statistical hypothesis that average weights
(\ref{avweight}) do not depend on phase - space variables and can
be calculated alone on basis of symmetry group consideration. This
also can be proved under the stronger assumption that in Eq.
(\ref{dsum}) any state with fixed $\Lambda, \lambda_\Lambda$ has
the same weight (see e.g. \cite{Cerul})

Let us consider a projection operator ${\cal P}^{\Lambda}$ on the
subspace spanned by all states transforming under representation
$\Lambda$.
\begin{equation} \label{proj1states}
{\cal P}^{\Lambda}\left\vert
N^{(\alpha_1)}_{\nu_{\alpha_1}},\dots,
N^{(\alpha_n)}_{\nu_{\alpha_n}}\right\rangle=\sum_{\xi_{\Lambda}}\!^{\oplus}
\left\vert \Lambda, \lambda_{\Lambda};\xi_\Lambda\right\rangle
 {\cal C}^{\Lambda,\lambda_{\Lambda}}_{\{N^{(\alpha_1)}_{\nu_{\alpha_1}},
\dots,\,N^{(\alpha_n)}_{\nu_{\alpha_n}}\}}(\xi_\Lambda)\,;
\end{equation}

This operator has the generic form (see e.g. \cite{Wigner}):
\begin{equation}\label{proj1}
{\cal
P}^{\Lambda}=d(\Lambda)\int\limits_G\,d\mu(g)\bar\chi^{(\Lambda)}(g)U(g)\,;
\end{equation}
Here $\chi^{(\Lambda)}$ is the character of the representation
$\Lambda$, $d\mu(g)$ is the invariant Haar measure on the group,
and $U(g)$ is an operator transforming a state under
consideration. We will use the matrix representation:
\begin{eqnarray}\label{transf}
& &U(g)\left\vert N^{(\alpha_1)}_{\nu_{\alpha_1}},\dots,
N^{(\alpha_n)}_{\nu_{\alpha_n}}\right\rangle \nonumber\\ &=
&\sum\limits_{\nu_1^{(1)},\dots,\nu_n^{(N_{\nu_n})}}\,
D^{(\alpha_1)}_{\nu_1^{(1)}\nu_1}\!\!\cdots
D^{(\alpha_1)}_{\nu_1^{(N_{\nu_1})}\nu_1}\!\!\cdots
D^{(\alpha_n)}_{\nu_n^{(1)}\nu_n}\!\!\cdots
D^{(\alpha_n)}_{\nu_n^{(N_{\nu_n})}\nu_n} \left\vert
N^{(\alpha_1)}_{\nu_{\alpha_1}},\dots,
N^{(\alpha_n)}_{\nu_{\alpha_n}}\right\rangle\ ;
\end{eqnarray}
$D^{(\alpha_n)}_{\nu,\nu}$ is a matrix elements of the group
element $g$ corresponding to the representation $\alpha$. Notation
convention in Eq.\,(\ref{transf}) arises since  there are
$N^{(\alpha_j)}_{\nu_{\alpha_j}}$ states transforming under
representation $\alpha_j$ and having quantum numbers of the
$\nu_{\alpha_j}$-th member of a given multiplet.

The statistical hypothesis identifies the average weight
$\overline{P^{\Lambda,\lambda_{\Lambda}}_{\{N^{(\alpha_1)}_{\nu_{\alpha_1}},
\dots,\,N^{(\alpha_n)}_{\nu_{\alpha_n}}\}}}$ with a norm of the
vector ${\cal P}^{\Lambda}\left\vert
N^{(\alpha_1)}_{\nu_{\alpha_1}},\dots,
N^{(\alpha_n)}_{\nu_{\alpha_n}}\right\rangle$. This norm can be
written as

\begin{eqnarray}\label{normP}
\left\langle N^{(\alpha_1)}_{\nu_{\alpha_1}},
\cdots,N^{(\alpha_n)}_{\nu_{\alpha_n}}\right\vert {\cal P
}^{\Lambda}\left\vert N^{(\alpha_1)}_{\nu_{\alpha_1}},\dots,
N^{(\alpha_n)}_{\nu_{\alpha_n}}\right\rangle =
\sum\limits_{\xi_\Lambda}\vert{\cal
C}^{\Lambda,\lambda_{\Lambda}}_{\{N^{(\alpha_1)}_{\nu_{\alpha_1}},
\dots,\,N^{(\alpha_n)}_{\nu_{\alpha_n}}\}}(\xi_\Lambda)\vert^2\,;
\end{eqnarray}

where the relation $({\cal P}^\Lambda)^2={\cal P}^\Lambda$ was
used.

Left hand side of this equation can be calculated directly from
Eqs.(\ref{proj1}) and (\ref{transf}). One gets finally
\begin{equation}
 \overline{P^{\Lambda,\lambda_{\Lambda}}_{\{N^{(\alpha_1)}_{\nu_{\alpha_1}},
\dots,\,N^{(\alpha_n)}_{\nu_{\alpha_n}}\}}}
  =
{\cal A}^{\{{\cal N}\}}
d(\Lambda)\int\limits_G\,d\mu(g)\bar\chi^{(\Lambda)}(g)
[D^{(\alpha_1)}_{\nu_1\nu_1}]^{N^{(\alpha_1)}_{\nu_{\alpha_1}}}\cdots
[D^{(\alpha_n)}_{\nu_n\nu_n}]^{N^{(\alpha_n)}_{\nu_{\alpha_n}}}\,;
\label{weights}
\end{equation}
where ${\cal A}^{\{{\cal N}\}}$ is a permutation normalization
factor. For particles of the kind $\{\alpha,\zeta\}$ we included
in Eq.\,(\ref{weights}) the permutation factor:
\begin{equation}\label{permfac1}
{\cal A}^\alpha_{(\zeta)}= \frac{{\cal
N}^{(\alpha)}_{(\zeta)}!}{\prod\limits_{\nu_\alpha} {\cal
N}^{(\alpha)}_{\nu_\alpha;(\zeta)}!}\,;
\end{equation}
The permutation factor ${\cal A}^{\{{\cal N}\}}$ is a product of
all "partial" factors
\begin{equation}\label{permfactor}
{\cal A}^{\{{\cal N}\}}= \prod\limits_j\prod_{\zeta_j}{\cal
A}^{\alpha_j}_{(\zeta_j)}\,;
\end{equation}
The permutation factor assures the normalization of state vectors:
\begin{equation}
\left\langle N^{(\alpha_1)}_{\nu_{\alpha_1}},
\cdots,N^{(\alpha_n)}_{\nu_{\alpha_n}}\right\vert \left.
N^{(\alpha_1)}_{\nu_{\alpha_1}},\dots,
N^{(\alpha_n)}_{\nu_{\alpha_n}}\right\rangle={\cal A}^{\{{\cal
N}\}}\ ; \label{norm}
\end{equation}
This normalization reflects an invariance of the state vector with
respect to permutations which shuffle indistinguishable particles.
\section{\normalsize\bf Incorporation of Symmetry}\label{sym}
\noindent The expression Eq.\,(\ref{weights}) is a starting point
for further considerations. It provides together with
Eq.\,(\ref{numb}) and
 Eq.\,(\ref{kin}) subsidiary constraints on distribution functions
$f^{(\alpha_i,\nu_i)}$. These conditions assure that in a
dynamical evolution the symmetry of the system is preserved. When
symmetry is conserved, then all weights in Eq.\,(\ref{weights})
are constant in time. In a case of strong interaction and colour
symmetry, all weights, except for the weight corresponding to the
singlet state, must remain zero.

We now convert the global constraint into a time evolution
condition and consider:
\begin{equation} \label{cond}
\frac{d}{dt}\overline{P^{\Lambda,\lambda_{\Lambda}}_{\{N^{(\alpha_1)}_{\nu_{\alpha_1}},
\dots,\,N^{(\alpha_n)}_{\nu_{\alpha_n}}\}}}=0\,;
\end{equation}
Introducing  here the result of Eq.\,(\ref{weights}) one obtains:
\begin{eqnarray} \label{deriv}
0&=&\frac{d\,{\cal A}^{\{{\cal N}\}}}{dt}
d(\Lambda)\int\limits_G\,d\mu(g)\bar\chi^{(\Lambda)}(g)
[D^{(\alpha_1)}_{\nu_1\nu_1}]^{N^{(\alpha_1)}_{\nu_{\alpha_1}}}\cdots
[D^{(\alpha_n)}_{\nu_n\nu_n}]^{N^{(\alpha_n)}_{\nu_{\alpha_n}}}
\nonumber\\ &\;+&\sum_{j=1}^n\sum_{\nu_{\alpha_j}}\,
\frac{d\,N^{(\alpha_j)}_{\nu_{\alpha_j}}}{dt} {\cal A}^{\{{\cal
N}\}}d(\Lambda) \int\limits_G\,d\mu(g)\bar\chi^{(\Lambda)}(g)
[D^{(\alpha_1)}_{\nu_1\nu_1}]^{N^{(\alpha_1)}_{\nu_{\alpha_1}}}\cdots
[D^{(\alpha_n)}_{\nu_n\nu_n}]^{N^{(\alpha_n)}_{\nu_{\alpha_n}}}
\log[D^{(\alpha_j)}_{\nu_j\nu_j}] \,;
\end{eqnarray}
All integrals which appear in Eq.\,(\ref{weights}) and
Eq.\,(\ref{deriv}) can be expressed explicitly in an analytic form
for any compact symmetry group.

To write an expression for the time derivative of the
normalization factor ${\cal A}^{\{{\cal N}\}}$ we perform analytic
continuation from integer to continuous values of variables
$N^{(\alpha_n)}_{\nu_{\alpha_n}}.$ Thus we replace all factorials
by the $\Gamma$--function  of corresponding arguments. We
encounter here also the digamma function $\psi$ \cite{Abram}:
\begin{equation}\label{digamma}
\psi(x)=\frac{d\, \log\Gamma(x)}{d\,x}\,;
\end{equation}
This allows to write for Eq.\,(\ref{deriv}):
\begin{equation}\label{evfactor}
\frac{d\,{\cal A}^{\{{\cal N}\}}}{dt} = {\cal A}^{\{{\cal
N}\}}\sum_j\sum_{\zeta_j}
 \left[\frac{d\,{\cal N}^{(\alpha_j)}_{(\zeta_j)}}{dt}
 \psi({\cal N}^{(\alpha_j)}_{(\zeta_j)}+1) -\sum_{\nu_{\alpha_j}}
\frac{d\,{\cal N}^{(\alpha_j)}_{\nu_{\alpha_j};(\zeta_j)}}{dt}
\psi({\cal N}^{(\alpha_j)}_{\nu_\alpha;(\zeta_j)}+1)\right]\,;
\end{equation}
To get a consistent analytical continuation in the number of
particles one should define the time derivatives ${d\,{\cal
N}^{(\alpha)}_{\nu_\alpha;(\zeta)}}/{dt}$. We define these  rates
of particle number change from the integrated Boltzmann kinetic
equation, Eq.\,(\ref{kin}), explicitly
\begin{eqnarray}\label{change}
\frac{d\,N^{(\alpha_i)}_{\nu_{\alpha_i}}}{dt}
&=&\sum_{\alpha_j,\alpha_k,\alpha_l}\sum_{\nu_j,\nu_k,\nu_l}
\sum_{\zeta_j,\zeta_k,\zeta_l} \int\,dV d\Gamma_j d\Gamma_k
d\Gamma_l d\Gamma_i\, {\cal
W}_{\nu_i\nu_j;\nu_k\nu_l}^{(\zeta_i,\zeta_j;\zeta_k,\zeta_l)}
(\Gamma_k,\Gamma_l;\Gamma_j,\Gamma_i)\nonumber\\ &\ & \left[ {\cal
F}^{(\alpha_i,\nu_i)}_{(\zeta_i)}(\Gamma_i,\vec r,t)
       {\cal F}^{(\alpha_j,\nu_j)}_{(\zeta_j)}(\Gamma_j,\vec r,t)
              f^{(\alpha_k,\nu_k)}_{(\zeta_k)}(\Gamma_k,\vec r,t)
              f^{(\alpha_l,\nu_l)}_{(\zeta_l)}(\Gamma_l,\vec r,t)
\right.\nonumber\\&&\left.
      -{\cal F}^{(\alpha_k,\nu_k)}_{(\zeta_k)}(\Gamma_k,\vec r,t)
       {\cal F}^{(\alpha_l,\nu_l)}_{(\zeta_l)}(\Gamma_l,\vec r,t)
              f^{(\alpha_i,\nu_i)}_{(\zeta_i)}(\Gamma_i,\vec r,t)
              f^{(\alpha_j,\nu_j)}_{(\zeta_j)}(\Gamma_j,\vec r,t)
\right]\,;
\end{eqnarray}
Contributions from gradient terms of Eq.\,(\ref{kin}) vanish due
to Gauss law. These terms are transformed  in surface integrals
and beyond the volume occupied by the system all distribution
functions are equal to zero.

Eqs\,(\ref{deriv},\ref{evfactor},\ref{change}) in fact constitute
the global subsidiary condition which should be fulfilled by the
 microscopic kinetic equations Eq.\,(\ref{kin}). These are the
necessary conditions for preserving the internal symmetry on the
macroscopic level. Rates of change
 ${d\,{\cal N}^{(\alpha)}_{\nu_\alpha;(\zeta)}}/{dt}$
are related to ``macrocurrents", which are counterparts of
``microcurrents" related directly to a symmetry on a microscopic
level via the Noether theorem. Eq.\,(\ref{change}) can be
considered as a set of conditions on macrocurrents to provide
consistency with the overall symmetry of the system. Therefore we
believe that this equation can also be used as a starting point
for multicomponent hydrodynamic equations with internal symmetry
properties taken into account. One should notice that this
subsidiary condition takes into account also surface effects for
the finite volume systems. This is due to the space variables
integration which is performed in Eqs\, (\ref{numb}) and
(\ref{change}).

One easily sees that for the case of abelian symmetry the two
constraints Eq.\,(\ref{weights}) and Eq.\,(\ref{deriv}) do not
lead to new results: first we recall that all irreducible
representations of abelian group are one-dimensional. Next, let
basic particles have ``charges" $q_1,\dots, q_n$\,, and let the
global charge be $Q$. Then the only consequence of
Eq.\,(\ref{weights}) follows for nonvanishing weight $Q=N_1
q_1+\cdots+N_n q_n$, which is a rather obvious result. New results
appear only for nonabelian symmetries.

\section{\normalsize\bf Example: Isospin}\label{isos}
\noindent
 We now consider as an example the case of the $SU(2)$ symmetry
with basic particles transforming under spinor $({\bf 1\over 2})$
(fundamental) and vector $(\bf 1)$ (adjoint) representations. This
example can be realized by a gas mixture of nucleons and pions. To
describe all group elements the three group's parameters
 $\alpha,\beta,\gamma$ are chosen in such a way that
diagonal matrix elements have the well known form \cite{Wigner}:\\
{\it i.)} for the fundamental representation $({\bf 1\over 2})$:
\begin{eqnarray}\label{j_half}
D^{(1/2)}_{mm}(\alpha,\beta,\gamma) =
e^{im(\alpha+\gamma)}\cos{\beta\over 2}\,;\qquad m=\pm{1\over
2}\,;
\end{eqnarray} {\it ii.)} and for the adjoint
representation $(\bf 1)$:\\
\begin{eqnarray}
D^{(1)}_{\pm 1,\pm 1}(\alpha,\beta,\gamma)
  &=&{1\over 2}\,e^{\pm im(\alpha+\gamma)}(1+\cos\beta)\,;\\
D^{(1)}_{0,0}(\alpha,\beta,\gamma) &=& \cos\beta\,;
\end{eqnarray}
The Haar measure for the $SU(2)$ group in this parametrization has
the form
\begin{eqnarray}
\int\,d\mu(g)f[g] ={1\over{8\pi^2}} \int\limits^{2\pi}_0d\alpha
\int\limits^{2\pi}_0d\gamma \int\limits^{\pi}_0d\beta\sin\beta
f[g(\alpha,\beta,\gamma)]\,;
\end{eqnarray}

Any  `macrostate' is made of an arbitrary number: $n_n,\, n_p,\,
n_{-},\, n_0, \, n_{+}$; where subscripts refer to members of the
fundamental representation  neutrons, protons; and members of
adjoint representations, $\pi^{-},\, \pi^0,\, \pi^{+}$,
 correspondingly.
Let us consider the special
 case when the macrostate is a $SU(2)$ singlet.
The weight of the singlet state is according to
Eq.\,(\ref{weights}):
\begin{eqnarray} \label{singlet}
& &\overline{P^{0,0}_{\{n_n,n_p,n_{-},n_0,n_{+}\}}}= \nonumber\\
&\ &{\cal A}^{\{{\cal N}\}}{1\over{8\pi^2}}
\int\limits^{2\pi}_0d\alpha\int\limits^{2\pi}_0d\gamma
\int\limits^{\pi}_0d\beta\sin\beta e^{-{i\over
2}(n_n-n_p+2n_{-}-2n_{+})(\alpha+\gamma)} \cos^{\cal R}{\beta\over
2}\cos^{n_0}\beta\nonumber\\ &&\qquad\qquad\qquad\qquad\equiv
{\cal A}^{\{{\cal N}\}}\widetilde
P^{0,0}_{\{n_n,n_p,n_{-},n_0,n_{+}\}}\,;
\end{eqnarray}
where
\begin{equation}\label{defR}
{\cal R}=n_n+n_p+2n_{-}+2n_{+}\,;
\end{equation}
The permutation normalization factor is here:
\begin{equation}\label{pifactor}
{\cal A}^{\{{\cal N}\}} =
\frac{(n_{-}+n_0+n_{+})!(n_n+n_p)!}{n_{-}!n_0!n_{+}!n_n!n_p!}\,;
\end{equation}
The real nonzero values of the weight is obtained only when the
argument of the exponent in  Eq.\.(\ref{singlet}) vanishes:
\begin{equation}\label{cond1}
n_n-n_p+2n_{-}-2n_{+}=0\,;
\end{equation}
This is equivalent to the conservation of the third component of
the isospin.

Novel behaviour is obtained only when one considers time evolution
of the system. Presence of an exact symmetry means that the
corresponding weight Eq.\,(\ref{weights}) is constant,  here we
consider the expression:
\begin{equation}\label{cond2}
\frac{d\ }{dt}\overline{P^{0,0}_{\{n_n,n_p,n_{-},n_0,n_{+}\}}}
 =0\,;
\end{equation}
We note that an appropriate analytical continuation should be
made. First we evaluate the integral appearing in
Eq.\,(\ref{singlet}):
\begin{equation}\label{singldiscr}
\frac{1}{{\cal A}^{\{{\cal
N}\}}}\overline{P^{0,0}_{\{n_n,n_p,n_{-},n_0,n_{+}\}}}=
(-1)^{n_0}\sum_{i=0}^{n_0}(-2)^i{n_0\choose i}\frac{1}{{\cal
R}+1+i}\,;
\end{equation}

This discrete form is not allowing an analytic continuation which
would allow for all necessary differentiations. However, we can
write the integral also as the hypergeometric ${}_2F_1$ function
\cite{Abram}:
\begin{equation} \label{hGauss}
\frac{1}{{\cal A}^{\{{\cal
N}\}}}\overline{P^{0,0}_{\{n_n,n_p,n_{-},n_0,n_{+}\}}} =
(-1)^{n_0}\frac{1}{{\cal R}+1} \, {}_2F_1(-n_0,{\cal R}+1,{\cal
R}+2;2)\,;
\end{equation}
where ${\cal R}$ is as defined in Eq.\,(\ref{defR}). We so obtain:
\begin{eqnarray} \label{singlcont}
&&\overline{P^{0,0}_{\{n_n,n_p,n_{-},n_0,n_{+}\}}}\nonumber\\
&&\;=\frac{\Gamma(n_{-}+n_0+n_{+}+1)
\Gamma(n_n+n_p+1)}{\Gamma(n_{-}+1)\Gamma(n_0+1)
\Gamma(n_{+}+1)\Gamma(n_n+1)\Gamma(n_p+1)} \frac{\cos
n_0\pi}{\Gamma(-n_0)}\sum_{i=0}^\infty\frac{\Gamma(-n_0+i)} {{\cal
R}+1+i}\frac{2^i}{i!}\,;
\end{eqnarray}
Eqs. (\ref{singlet}), (\ref{cond2}) and (\ref{singlcont}),
together with the condition (\ref{cond1})  result in:
\begin{eqnarray}\label{cond3}
0&=&2\frac{\partial \widetilde
P^{0,0}_{\{n_n,n_p,n_{-},n_0,n_{+}\}}}{\partial{\cal R}}
\left(\frac{d n_n}{dt} + 2\frac{d n_{-}}{dt}\right)+
\frac{\partial\widetilde
P^{0,0}_{\{n_n,n_p,n_{-},n_0,n_{+}\}}}{\partial n_0} \frac{d
n_0}{dt}\nonumber\\ &+&\frac{d\,\log{\cal A}^{({\cal
N})}}{dt}\widetilde P^{0,0}_{\{n_n,n_p,n_{-},n_0,n_{+}\}} \,;
\end{eqnarray}
The projection  integrals determine the coefficients which are,
explicitly:
\begin{eqnarray}
\frac{\partial \widetilde
P^{0,0}_{\{n_n,n_p,n_{-},n_0,n_{+}\}}}{\partial n_0}&=&
-\pi\frac{\sin
n_0\pi}{\Gamma(-n_0)}\sum_{i=0}^\infty\frac{\Gamma(-n_0+i)} {{\cal
R}+1+i}\frac{2^i}{i!} \nonumber\\& &+ \frac{\cos
n_0\pi}{\Gamma(-n_0)}\sum_{i=0}^\infty\frac{\Gamma(-n_0+i)
[\psi(-n_0)-\psi(-n_0+i)]}{{\cal R}+1+i}\frac{2^i}{i!}\\ &=&
(-1)^{n_0}\sum_{i=1}^{n_0}{n_0\choose i} \frac{
[\psi(1+n_0)-\psi(1+n_0-i)]}{{\cal R}+1+i}(-2)^i\,;\nonumber
\\
\frac{\partial\widetilde
P^{0,0}_{\{n_n,n_p,n_{-},n_0,n_{+}\}}}{\partial{\cal R}} &=&
(-1)^{n_0+1}\sum_{i=0}^{n_0}(-2)^i {n_0\choose i}\frac{1}{({\cal
R}+1+i)^2}\,;
\end{eqnarray}
and
\begin{eqnarray}\label{factperm}
\frac{d\,\log{\cal A}^{({\cal N})}}{dt}
&=&\frac{dn_n}{dt}\left[\psi(n_N+1)-\psi(n_n+1)\right]
+\frac{dn_p}{dt}\left[\psi(n_N+1)-\psi(n_p+1)\right]\nonumber\\
&+&\frac{dn_{-}}{dt}\left[\psi(n_\pi+1)-\psi(n_{-}+1)\right]+
\frac{dn_0}{dt}\left[\psi(n_\pi+1)-\psi(n_0+1)\right]\nonumber\\
&+&\frac{dn_{+}}{dt}\left[\psi(n_\pi+1)-\psi(n_{+}+1)\right]\,;
\end{eqnarray}
where $n_N=n_n+n_p$ is the total number of nucleons and
$n_\pi=n_{-}+n_0+n_{+}$ is the total number of pions.
Eq.\,(\ref{factperm}) can be also written in the form:

\begin{eqnarray}\label{factperm2}
\frac{d\,\log{\cal A}^{({\cal N})}}{dt}
&=&\frac{dn_n}{dt}\sum^{n_N}_{k=1+n_n}{1\over k}
+\frac{dn_p}{dt}\sum^{n_N}_{k=1+n_p}{1\over k}\nonumber\\
&+&\frac{dn_{-}}{dt}\sum^{n_\pi}_{k=1+n_{-}}{1\over k}+
\frac{dn_0}{dt}\sum^{n_\pi}_{k=1+n_0}{1\over k}
+\frac{dn_{+}}{dt}\sum^{n_\pi}_{k=1+n_{+}}{1\over k}\,;
\end{eqnarray}

Eqs.\,(\ref{cond3}--\ref{factperm2}) offer the final result for
the $SU(2)$ case . Notably, they imply a  relation for the number
of neutral pions in a system. We thus see, that when the case of
the non-abelian symmetry is carefully considered and not ignored,
one can get relations determining also the `neutral' members of
multiplets. In the standard approach the multiplicity of neutral
particles is obtained by introducing a subsidiary chemical
potential which is related to the lack of chemical equilibrium of
a system \cite{Kataja} or to the residual interaction with the
environment \cite{Turko2}.

\section{\normalsize\bf Conclusions and Outlook}\label{conc}
\noindent
 We have shown how constraints due to the preservation of the
symmetry properties of a multiparticle macroscopic state define
the path of evolution of the system. One should notice that
results we presented are general and do not depend on the
particular choice of the representation of the symmetry group. For
different initial symmetry group representations one gets
different paths, but they are all of a similar ``shape",
considering the hypothesis that the global behaviour of a
macrosystem (in a sense of statistical physics) should not be
altered  if a number of particles is changed by a very small
(``microscopic") amount.

We have explicitly presented the example how our constraint works
in the simplest non - trivial case of $SU(2)$ symmetry group.

Although we have studied and implemented the discreet symmetry
using quantum states, whenever we referred here to a dynamical
equation we considered the limit of incoherent state evolution
described by the Boltzmann equation. The dynamical evolution we
consider thus is described in terms of diagonal density matrix.
This is the appropriate approach given that our main objective is
to arrive at a dynamical derivation of symmetry deformed
statistical distribution.

We recall  that quantum correlations (without symmetry) alone are
responsible for the deformation of the Boltzmann distribution into
Bose/Fermi distributions, and that the Boltzmann equation yields
this result when we allow for Fermi blocking/Bose enhancement in
the collision term. In that line of thought, the next step would
be to show that it is possible to obtain now within a dynamical
Boltzmann equation calculation the evolution of a many body system
into symmetry-deformed statistical equilibrium distribution. We
are also exploring the possibility that the methods here presented
allow the formulation of a microscopic transport theory which
would obey the long range correlations introduced by the
macroscopic quantum and symmetry constraints.


\section*{\normalsize\it Acknowledgments}\noindent
Work supported in part by a grant from the U.S. Department of Energy,
DE-FG03-95ER40937\,, and in part by the Polish Committee for
Scientific Research under contract KBN - 2 P03B 030 18\,.

\end{document}